\documentclass[journal=jctcce,manuscript=article]{achemso}

\author{Ivan M. Zeron}
\affiliation{Departamento de Qu\'{\i}mica F\'{\i}sica, Facultad de Ciencias
Qu\'{\i}micas, Universidad Complutense de Madrid, 28040 Madrid, Spain}
\altaffiliation{Contributed equally to this work}
\author{Miguel A. Gonzalez}
\affiliation{Departamento de Qu\'{\i}mica F\'{\i}sica, Facultad de Ciencias
Qu\'{\i}micas, Universidad Complutense de Madrid, 28040 Madrid, Spain}
\altaffiliation{Contributed equally to this work}
\alsoaffiliation{Current address: Universidad Rey Juan Carlos, ESCET, 28933
Mostoles, Madrid, Spain}
\author{Edoardo Errani} 
\affiliation{Departamento de Qu\'{\i}mica F\'{\i}sica, Facultad de Ciencias
Qu\'{\i}micas, Universidad Complutense de Madrid, 28040 Madrid, Spain}
\author{Carlos Vega}
\affiliation{Departamento de Qu\'{\i}mica F\'{\i}sica, Facultad de Ciencias
Qu\'{\i}micas, Universidad Complutense de Madrid, 28040 Madrid, Spain}
\author{Jose L. F. Abascal}
\affiliation{Departamento de Qu\'{\i}mica F\'{\i}sica, Facultad de Ciencias
Qu\'{\i}micas, Universidad Complutense de Madrid, 28040 Madrid, Spain}
\email{abascal@ucm.es}

\title{``In Silico'' Seawater}

\begin{document}

\def\deg{$^{\rm o}$C}

\begin{tocentry}
\includegraphics*[clip,scale=0.93]{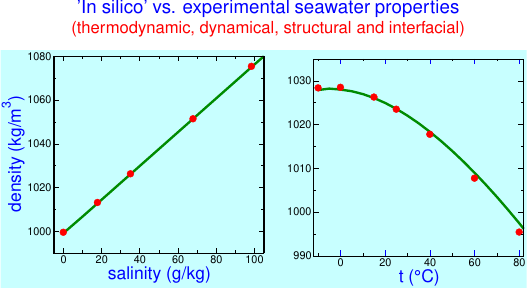}
\end{tocentry}

\begin{abstract}
Many important processes affecting the Earth's climate are determined by the physical properties of seawater. In addition, desalination of seawater is a significant source of drinking water for the human population living in coastal areas.
Since the physical properties of seawater governing these processes depend on the molecular interactions among its components, a deeper knowledge of seawater at the molecular level would contribute to a better understanding of these phenomena. However, in strong contrast with the situation in other areas such as biomolecules or materials science, molecular simulation studies reporting the physical properties of seawater are currently lacking. This is probably due to the usual perception of the seawater composition being too complex to approach.
This point of view ignores the fact that physical properties of seawater are dependent on a single parameter representing the composition, namely the salinity. This is because the relative proportions of any two major constituents of seasalt are always the same.
Another obstacle to performing molecular simulations of seawater could have been the unavailability of a satisfactory force field representing the interactions between water molecules and dissolved substances. However this drawback has recently been overcome with the proposal of the Madrid-2019 force field. 
In this work we show for the first time that molecular simulation of seawater is feasible. We have performed molecular dynamics simulations of a system, the composition of which is close to the average composition of standard seawater and with the molecular interactions given by the Madrid-2019 force field.
In this way we are able to provide quantitative or semiquantitative predictions for a number of relevant physical properties of seawater for temperatures and salinities from the oceanographic range to those relevant to desalination processes. The computed magnitudes include static (density), dynamical (viscosity and diffusion coefficients), structural (ionic hydration, ion-ion distribution functions) and interfacial (surface tension) properties.
\end{abstract}

\section{Introduction}
%
Seawater is a complex solution of substances, mostly ions, in water. 
The complex composition of seawater is often perceived as a barrier to
carry out numerical studies of its properties at the molecular level.
In fact, molecular simulation studies reporting  the physical properties of
seawater are currently lacking.
This is in strong contrast with the situation in other areas like biomolecules
or materials science. While in these areas molecular simulation now plays
an essential role as a complementary technique to experimental
measurements, similar applications to investigate features of salty systems of
difficult experimental access is much more limited.
However, it is important to note that, despite the complexity of the seasalt
composition, the most relevant physical properties of marine water ---leaving
aside surface or coastal effects--- depend essentially on a single parameter
representing the composition, namely the salinity.
Salinity is then a fundamental property of seawater and basic to understanding
biological and physical processes in oceans.
The absolute salinity is defined as the total amount of dissolved substances
(in grams) per kilogram of seawater\cite{unesco85}. The oceans salinity is
usually between 31 and 38 g/kg but higher ranges are relevant to desalination
processes.

The apparent contradiction between the complexity of seasalt composition and
the fact that seawater properties are essentially dependent on the salinity can
be easily explained. 
Irrespective of the total salinity, the relative proportions of any two major
constituents of seasalt are always the same. This evidence is known from the
beginning of the nineteenth century\cite{marcet19,forchhammer62} and it is
sometimes referred to as the Principle of Constant Proportions.
The Principle of Constant Proportions allows the definition of a precise reference
composition for Standard Seawater. The latter (arbitrary) definition was first
proposed by Knudsen\cite{knudsen03} and refers to certain surface seawater
samples taken from the North Atlantic ocean.
Currently, the International Association for the Physical Sciences of the
oceans (IAPSO) oversees the preparation of Standard Seawater to ensure the
quality and comparability of salinity data worldwide.
In what follows we will simply refer to IAPSO Standard Seawater as seawater.
In 2008, a Reference Composition ---consisting of the proportions of components
of seawater--- was defined.\cite{millero08}
Salinity differences are then caused by either evaporating fresh water or
adding fresh water from rivers and melted ice. In some way, this allows us to
consider seawater as a two component mixture, seasalt and water, the salinity
being the representative variable for the solution concentration.

The temperature of the oceans shows a characteristic vertical profile, called the
thermocline, which depends on the latitude. Similar patterns are found for the
salinity, the halocline. Since temperature and salinity determine the
thermodynamic properties of seawater, density exhibits the same type of
vertical profile as the thermocline and halocline.
This means that seawater does not mix vertically. The mixing is produced by
ocean currents, as shown for the first time by Sandstrom in
1908.\cite{sandstrom08}
These currents give rise to the thermohaline circulation\cite{rahmstorf03}
which plays a decisive role on the Earth's climate\cite{rahmstorf02} and on the
marine biology.
In this way, knowledge of the dependence of the density on the temperature and
salinity of marine water is crucial for the understanding and modeling of
the thermohaline circulation.

Thermophysical properties of seawater are well known. In fact, there is a Gibbs
function formulation, denoted as TEOS-10\cite{TEOS10} from which all the
thermodynamic properties of seawater can be consistently derived.
Apart of the limited range of the variables concerned (temperature, pressure
and salinity) which are extended in other thermodynamics
formulations\cite{sharqawy10} these equations are quite complex and do not allow
greater understanding of the role of the equation of state in the setting of
the large-scale circulation.
Although there are attempts to define simplified yet realistic equations of
state for seawater\cite{roquet15}, we believe that molecular simulation is the
ideal tool to get a better understanding of the thermodynamic properties
of seawater.

Molecular simulation\cite{frenkel96} has been recognized as a powerful tool to
investigate the properties of molecular systems. It is complementary to
experimental measurements and may be of great help in conditions under which
performing an experiment is a challenging
task.\cite{debenedetti20,corradini10,zaragoza20} However, to the best
of our knowledge, it has never been used to investigate seawater. There are two
likely reasons for that.
First, it might seem that the complex seawater composition and its dependence
on latitude and depth would prevent an investigation of global interest.
Nevertheless, this point of view does not take into account the relative
simplicity derived from the principle of constant proportions.
Here we show that a system containing a reduced number of ions mimics very
acceptably the mole fractions of seasalt as defined by the seawater Reference
Composition and that the resulting system with added water is suitable for
computer simulation.

On the other hand, molecular simulation relies on the availability of a good
description of the interactions between ions and water.
In recent years we have seen important advances regarding this 
topic.\cite{orozco14,jiang15,nezbeda16,smith18,baranyai20,panagiotopoulos20}
As we will see below, the complexity of the seawater composition implies the
simulation of considerably large samples. In these conditions, the use of
polarizable models would require huge computational resources so we focus our
interest on a rigid nonpolarizable model for water and molecular ions.
A very fruitful idea has been the incorporation of the electronic continuum
correction\cite{leontyev09}. It has been argued that polarizable models can be
reduced to nonpolarizable equivalent models with scaled charges.\cite{leontyev11}
Force fields based on this idea yielded an unprecedented agreement with the
experimental properties of electrolyte
solutions.\cite{pluharova13,kirby19,kann14,lebreton20}
In 2017 we proposed a force field for NaCl in water\cite{benavides17b} based on
the widely tested TIP4P/2005\cite{abascal05b,vega11} water model and the
use of scaled charges for the ions. The success of this work prompted us to
extend it and develop a force field, termed as Madrid-2019, including
parameters for the more abundant ions of seawater.\cite{zeron19}. We are thus
in a good position to address the issue of investigating the properties of
seawater from a molecular perspective.

In this work we study, using molecular simulation with the Madrid-2019
force field, the equation of state of seawater, i.e., the dependence of density
on temperature and salinity, $\rho=\rho(T,S)$.
We also check the performance of the force field for other magnitudes
representative of the thermophysical behavior of seawater.
In particular, we have also evaluated dynamical (viscosity and diffusion
coefficients), structural (ionic hydration, ion-ion distribution functions) and
interfacial (surface tension) properties. It will be shown that the
thermophysical properties of seawater can be satisfactorily predicted by
molecular simulations with the Madrid-2019 force field.

\section{``In Silico'' seasalt}

The Reference Composition of Standard Seawater is {\em defined}\cite{millero08}
in terms of the mole fractions of the solute components with a precision of
1/10$^7$. The number of water molecules required to produce a seawater sample
with the Reference Composition and a salinity around 35~g/kg would then be of
the order of $5\times 10^8$.
These numbers are obviously not adequate for computer simulation. It is also
arguable that very minor components cannot significantly affect the physical
properties of seawater. One of the goals of this work is to propose a 
simplified composition (which we term as ``in silico'' seasalt, ISSS) that 
enables a trustworthy comparison between simulation and experiment.

The first step is to shorten the list of components of seawater. For example,
the simulation of a salty solution mimicking as close as possible the seasalt
composition but containing just one CO$_{\rm 2}$ and one OH$^{\rm -}$ anion
would still require around $7\times10^7$ water molecules.
The six most abundant ions ---chloride, sodium, sulfate, magnesium, calcium,
and potassium--- represent 99.7 percent of the seasalt
mole fraction. The addition of two more ions to the list ---bicarbonate
and bromide--- would elevate the percentage to 99.9\%.
It is important to note that constituents such as dissolved gases (CO$_{\rm 2}$,
 O$_{\rm 2}$) or inorganic salts (made of phosphorus or nitrogen) may play an
essential role in climate regulation or in biological productivity but are
irrelevant regarding the physical properties.
In fact, neither O$_{\rm 2}$ nor any phosphorus or nitrogen compound form a part
of the Reference Composition definition of seawater.
A solution containing bicarbonate and bromide could indeed be tractable with
current computer resources. However, the interaction potentials for these ions are
presently much less reliable than those for the most abundant ions. We have thus
decided at this stage to describe our ``in silico'' seasalt by including
explicitly only the six ions mentioned above.

From the mole fractions and the ionic charge of the six components mentioned it
is clear that the charge balance is not easily satisfied. This is 
related to the fact that most of the neglected constituents carry a negative
charge. It is then convenient to include also a special type of solute by
grouping together the mole fraction and charge of the minor components.
This ``minor components'' solute would carry a negative charge and its assigned
molecular mass would be that required for the average molecular mass of the
ionic sample to be the same as that of the Reference Composition, namely
31.4038218 g/mol.\cite{millero08}

We have carried out a systematic search (see Supporting information)
looking for optimal sample sizes that reproduce as close as possible the
Reference Composition with a minimum of solute molecules.
Among them, we have arrived at a relatively small sample containing six types
of ions and a total of 318 solute molecules (see Table~\ref{tab:ISSW}).
The sample, which will be referred to as ISSS, reproduces the
Reference Composition mole fractions with a root mean square deviation around
0.0003. Table~\ref{tab:ISSW} shows the number of ions of each type and
compares the molality of the components with the corresponding Reference
Composition definition. It also shows the amounts of each component in the
Reference Composition Standard Seawater and in the ISSS+water (318 ions +15210
water molecules) systems at $S=35.165$~g/kg salinity.
In the Supporting Information we give a detailed description of a number of optimal
compositions for up to 13 seasalt components.
%
\begin{table*}
\centering
\caption{
Composition of the ``in silico'' seasalt, ISSS, used in this work. In order to
satisfy the electroneutrality of the solution, the charge of the ``minor
components'' is the same as that of the chloride anions.  Second column shows
the mole fractions defined in the Reference Composition for solutes with
$x>0.002$\cite{millero08}. Third column displays the amounts (g) of each
component in the Reference Seawater at $S=35.16504$~g/kg salinity.  Fourth and
fifth columns give the number of ions in the ISSS sample and their
corresponding mole fractions. The last column presents the amounts (g) of each
component in the ISSS+water system at $S=35.165$~g/kg.  A value of 63 g/mol
is assumed for the molecular mass of the ``minor components''.  With this
choice, the average molecular mass of the ISSS seasalt is 31.404 g/mol, the
same as that of the Reference Composition of Standard Seawater.}
\label{tab:ISSW}
\begin{tabular}{|l | c c | c c c|}
\hline
Component   &$x_{ion}$
		       &   mass   &$n_{ISSS}$
                                         &$x_{ISSS}$&$m_{ISSS+water}$\\
\hline
 Chloride  & 0.4874839 & 19.35271 &   155 & 0.4874 & 19.349 \\
 Sodium    & 0.4188071 & 10.78145 &   133 & 0.4182 & 10.766 \\
 Sulfate   & 0.0252152 &  2.71235 &     8 & 0.0252 &  2.706 \\
 Magnesium & 0.0471678 &  1.28372 &    15 & 0.0472 &  1.284 \\
 Calcium   & 0.0091823 &  0.41208 &     3 & 0.0094 &  0.423 \\
 Potassium & 0.0091159 &  0.39910 &     3 & 0.0094 &  0.413 \\
 Minor components                    
           & 0.0030278 &  0.22363 &     1 & 0.00314&  0.222 \\
 Sum       & 1.0000000 & 35.16504 &   318 & 1.0000 & 35.164 \\
\hline
 Water     &           & 964.83496& 15210 &        & 964.836 \\
\hline
\end{tabular}
\end{table*}

\section{Methods}
In this work we use molecular dynamics simulation to investigate the properties
of a system mimicking the composition of Standard Seawater.  We have
chosen the successful TIP4P/2005 model\cite{abascal05b} to account for the
interactions between the water molecules. For the ion-ion and ion-water
interactions we have used a recent parametrization\cite{zeron19} ---the
so-called Madrid-2019 force field--- based on the use of scaled charges and
Lennard-Jones (LJ) short range interactions for the ions.
It includes parameters for all the ionic species present in our ``in silico''
seawater model (for simplicity we also  assigned to the minor component the LJ
parameters of the chloride ion).
                Ion-water and the more important ion-ion cross interactions
were explicitly optimized. For the rest of the ion-ion cross interactions, the 
Lorentz-Berthelot combining rules were employed.
Most of the systems are made of 15210 water molecules and the number of ions
are multiples of the ISSS sample shown in Table~\ref{tab:ISSW}. We have also
studied a system at a lower salinity by considering the 318 ions of the ISSS
sample and increasing the number of water molecules to 30420.

The simulations have been performed using GROMACS 4.5.5\cite{hess08} with a
2~fs time step. The cutoff radii have been set to 0.95~nm for the
Lennard-Jones interactions. Long range corrections to the
Lennard-Jones potential energy and pressure were included. Long range
electrostatic interactions have been evaluated with the smooth Particle Mesh
Ewald method.\cite{frenkel96} The geometry of the water molecules and of the
sulfate ions has been enforced using {\em ad hoc} constraints, in particular,
the SHAKE algorithm. The Nos\'e-Hoover thermostat has been applied to set the
temperatures at the desired values. All the simulations in the
isobaric-isothermal (NpT) ensemble were performed at a fixed pressure of 1~bar
by means of an isotropic Parrinello-Rahman barostat. 
The average volumes obtained in the NpT runs were used as input
for the simulations at constant volume performed to evaluate the viscosities.
The simulated time of the runs has been typically 10-20~ns for the calculation
of the water-ions structure and residence times and, at least, 100~ns for the
evaluation of viscosities, surface tension, ion diffusivities and ion-ion rdf's.
Exceptionally a run of 500~ns has been carried out for the state at 15\deg,
$S$= 35.165 g/kg in order to increase the accuracy of the diffusivity of the
minor components.

The evaluation of some properties such as the density or the radial
distribution function from a molecular dynamics run is trivial and may be
found in textbooks.\cite{frenkel96}
However, the determination of other quantities still requires some
methodological comments.
For the evaluation of the viscosity we have used the Green-Kubo formula
\begin{equation}
\label{shear_GK}
 \eta = \frac {V}{kT} \int_{0}^{\infty} 
 \langle P_{\alpha\beta}(t_{0})\; P_{\alpha\beta}(t_{0}+t) \rangle_{t_{0}}\; dt,
\end{equation}
where $P_{\alpha\beta}(t)$ is a component of the pressure tension and the
brackets denote the ensemble average.
Some care is needed\cite{gonzalez10,zhang15} to select
the upper limit and the asymptotic value of the integral.
The self-diffusion coefficients have been evaluated by means of the Einstein
relation
\begin{equation}
\label{eq_diffus}
  D = \lim_{t \to \infty} \frac{1}{6 t} \langle 
	[\mathbf{r}_{i}(t)-\mathbf{r}_{i}(t_{0})]^{2} \rangle ,
\end{equation}
where $\mathbf{r}_{i}(t)$ and $\mathbf{r}_{i}(t_{0})$ are the positions of the
i$-th$ particle at time t and a certain origin of time t$_{0}$.
For the calculation of the surface tension $\gamma$ we considered a slab of
liquid placed between two empty regions. In the case of a planar interface,
$\gamma$ is evaluated as
\begin{equation}
 \gamma = \frac{L_z}{2} ({\bar p_N} - {\bar p_T}),
\end{equation}
where $\bar p_N$ and $\bar p_T$ are the macroscopic normal and tangential
components of the pressure tensor and $L_z$ is the length of the simulation
box along the direction perpendicular to the interface.

It is well known that the truncation of the potential significantly affects
the interfacial properties.\cite{trokhymchuk99}
For the calculation of $\gamma$ we have extended the cutoff radii $r_c$.
At several salinities 
we have performed two sets of calculations using $r_c=1.3$~nm and $r_c=1.6$~nm
and we have observed that the difference between the surface tension at a given
salinity with respect to that for pure water $\gamma(S)$-$\gamma(0)$ is
independent of the cutoff radii provided that both runs use the same cutoff.
The values reported in this work correspond to simulations using $r_c=1.3$~nm.
The self-diffusion coefficients are also quite sensitive to finite-size
effects\cite{yeh04,celebi20}
but the size of our samples is already large enough to avoid this problem.
In addition, it is expected that the error should cancel when we compare two
values under identical conditions. Thus, $D(S)/D(0)$ is a more appropriate 
magnitude to investigate the diffusivity in seawater than $D(S)$ alone.

The hydration numbers, HN, are the average number of water molecules in the
first solvation shell of an ion and are trivially obtained by integrating
the corresponding water-ion rdf's. The residence times of these water molecules
may be calculated from time correlation functions\cite{koneshan98}
\begin{equation}
	R(t) = \frac{1}{N_h} \sum_{i=1}^{N_h} [\Theta_i(0) \Theta_i(t)],
\end{equation}
where $\Theta_i(t)$ is the Heaviside unit step function, which is 1 if a
water molecule $i$ is in the coordination shell of the ion at time
$t$ and zero otherwise, and $N_h$ is the hydration number of this
shell. From $R(t)$
it is possible to evaluate an average lifetime, $\tau$, defined as
\begin{equation}
       \tau = \int_{0}^{\infty} \left < R(t) \right > dt.
\end{equation}
Typically, $R(t)$ exhibits an exponential decay at short times,
$R(t)\propto\exp(-t/\tau)$ so it can also be estimated from the
corresponding fit. 

The uncertainties of most of the quantities (with the exception of the
viscosity) have been estimated from block averages\cite{hess02,gonzalez16}.
The estimated error of the densities is very small, usually $<0.02$\% because
the density is in most cases a byproduct of the calculation of the viscosity or
the diffusion coefficients. 
The uncertainty of the viscosities has been estimated as the standard deviation
of the results in the complete runs for each of the five independent pressure
components.\cite{gonzalez10}

\section{Results}
%
\subsection{Results for the Madrid-2019 force field}
\begin{figure}[tbhp]
\caption{Comparison of the densities calculated in this work (points) with
experimental data.\cite{sharqawy10} Left: as a function of salinity
for seawater at 15\deg; right: as a function of temperature for seawater at
$S$=35.165 g/kg. The uncertainties are in all cases much smaller than the point
size.}
\centering
\includegraphics*[clip,scale=0.40]{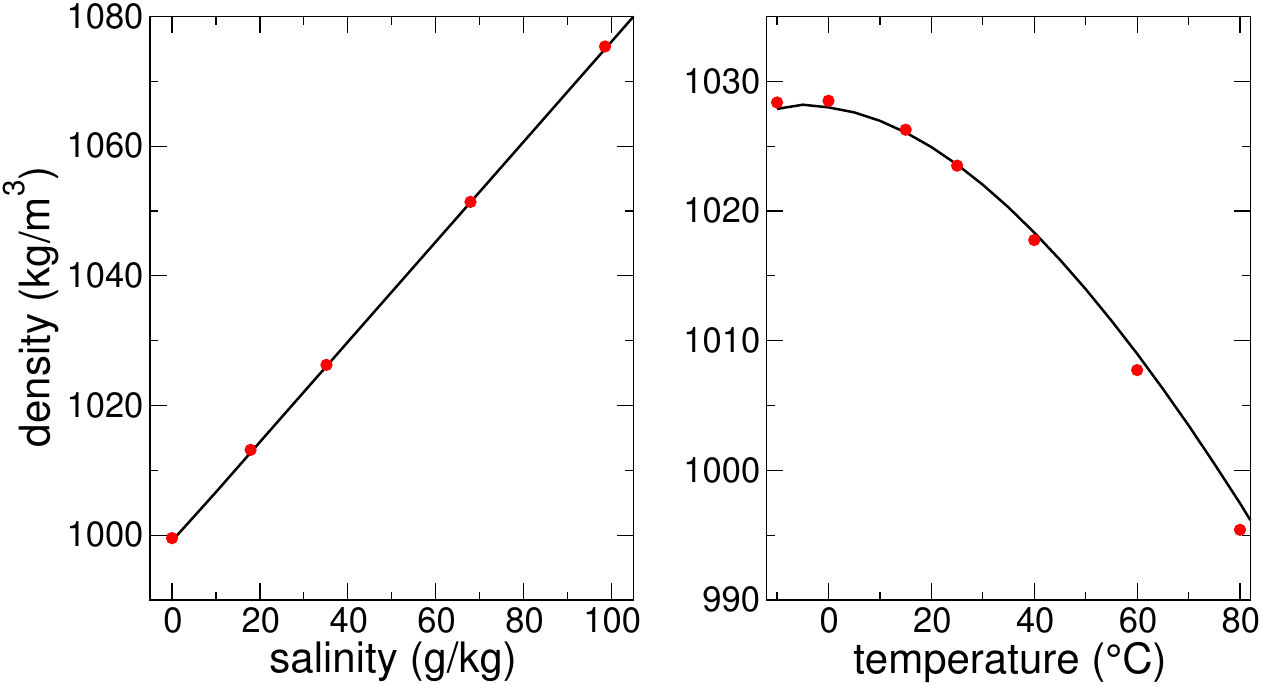}
\label{fig:dens}
\end{figure}

In this section we compare the molecular dynamics results for the ISSS+water
system with experimental measurements of Standard Seawater.
Fig.~\ref{fig:dens} shows that the calculated densities essentially match the
experimental data and its dependence on salinity at 15\deg\ (close to
the average temperature of the ocean surface waters). Notice that the
salinity of the more concentrated solution in Fig.~\ref{fig:dens} is about
three times larger than the average salinity of open seas. As for the dependence
of the density on temperature we may observe in Fig.~\ref{fig:dens} that the
predictions of the ``in silico'' system are quite accurate for $S$=35.165~g/kg.
The simulation results show an excellent performance up to 40\deg\ and the
agreement between simulation and experiment degrades slightly at higher
temperatures.
%

\begin{center}
\begin{figure}[tbhp]
\caption{Comparison of the viscosities calculated in this work (points) with
experimental data.\cite{sharqawy10} Left: as a function of salinity
for seawater at 15\deg; right: as a function of temperature for seawater at
$S$=35.165 g/kg}
\centering
\includegraphics*[clip,scale=0.40]{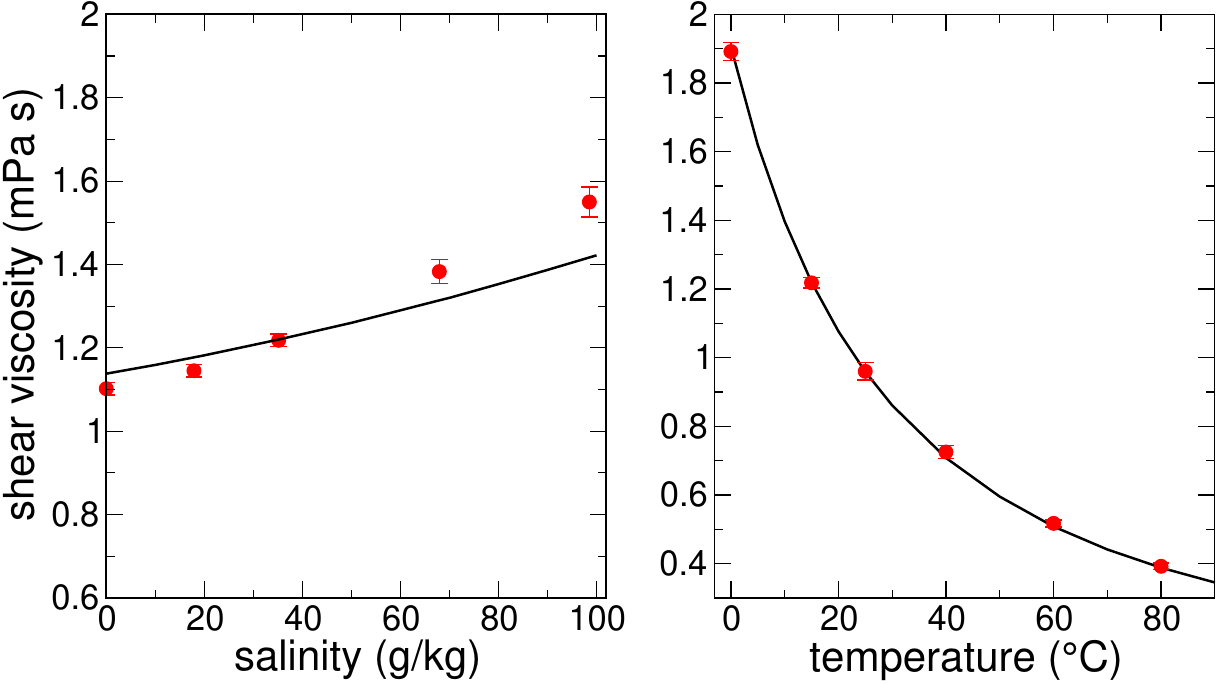}
\label{fig:visco}
\end{figure}
\end{center}

The results for the shear viscosity are presented in Fig.~\ref{fig:visco}.
The simulation data reproduce adequately the dependence of the viscosity on the
salinity though the slope of the curve is a bit steeper
than that of the experimental values. Interestingly, the better agreement is
found just for the more relevant seawater salinities region around 35~g/kg
where experiment and numerical predictions are almost coincident.
Fig.~\ref{fig:visco} shows that the agreement extends over a wide range of
temperatures. It is a fortunate coincidence that in addition to a generally
excellent performance of the model, the better predictions correspond to
temperatures and salinities around the average values of seawater in oceans. 

\begin{center}
\begin{figure*}[!ht]
\caption{Top left: Self-diffusion coefficients of water as a function of the
inverse of the temperature for ``in silico'' seawater at $S$=35.165 g/kg in
logarithmic scale (red circles). The blue symbols represent the ratio between
the simulation values of the diffusion coefficients in seawater and those in
simulations of pure water at the same temperature.
Top right: Simulation values for $D_w$, $\eta$, and $\eta\;D_w$ as a function of
temperature at $S$=35.165 g/kg (units as in previous figures).
Bottom: Self-diffusion coefficients of water as a function of salinity for
seawater at 15\deg.}
\centering
\includegraphics*[clip,scale=0.57]{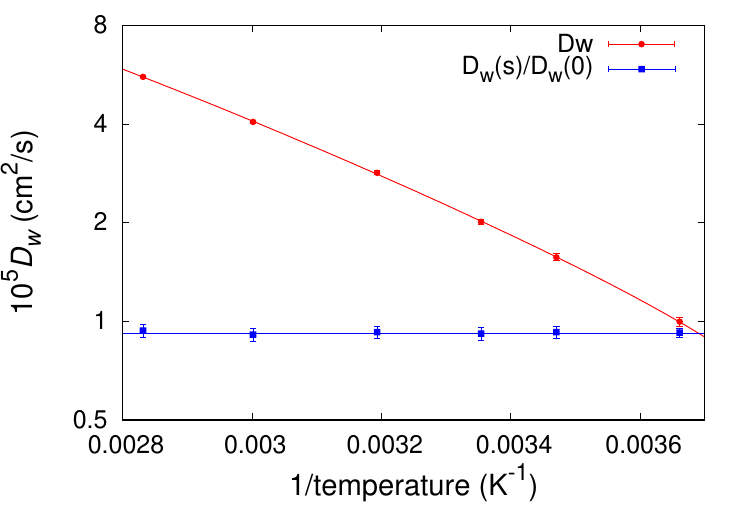}
\includegraphics*[clip,scale=0.57]{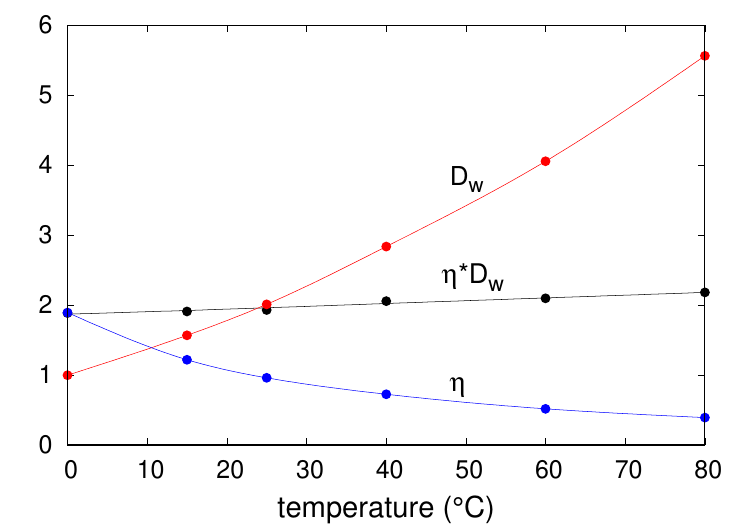} \\
\includegraphics*[clip,scale=0.57]{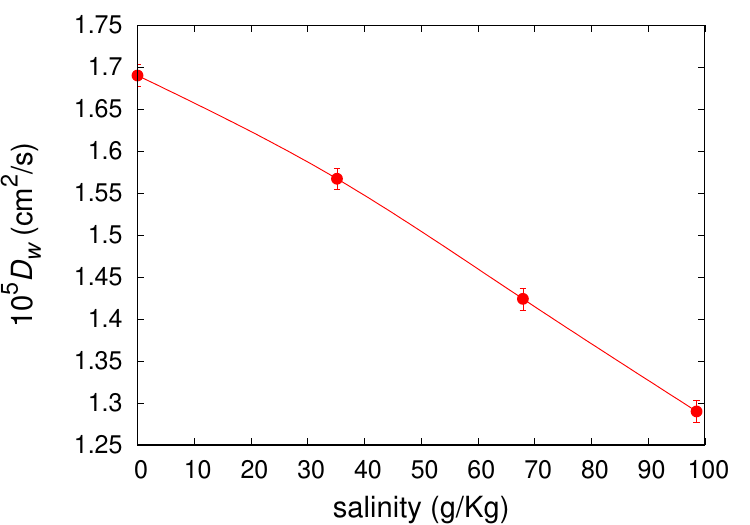}
\label{fig:Dw}
\end{figure*}
\end{center}

\begin{figure}[!ht]
\caption{Self-diffusion coefficients of cations (left) and anions (right) as a
function of the inverse of the temperature for seawater at $S$=35.165 g/kg in
logarithmic scale.
Small symbols are the simulation results and large symbols are the experimental
data.\cite{li74,poisson83} Lines are a quadratic fit of the numerical results.}
\centering
\includegraphics*[clip,scale=0.41]{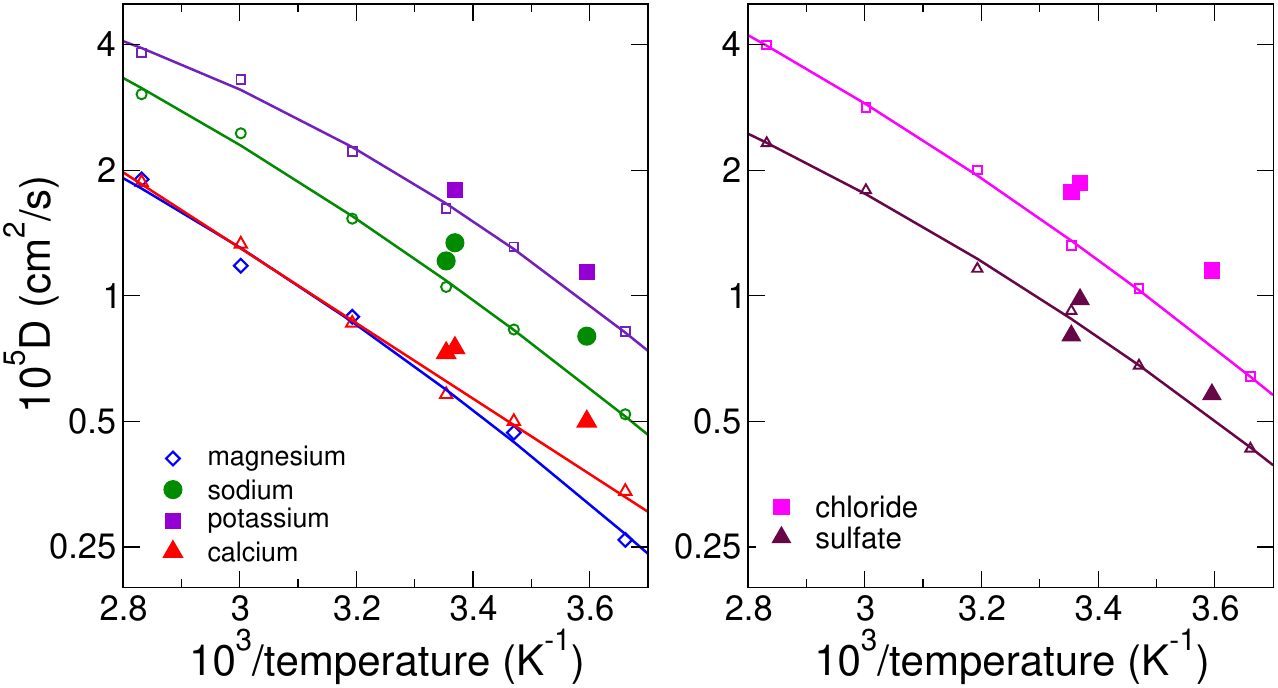}
\label{fig:Dions}
\end{figure}

The self-diffusion coefficient of the water molecules, $D_w$, in seawater of
salinity $S=35.165$~g/kg exhibits an almost perfect Arrhenius behavior in the 
range of temperatures from 0 to 80\deg\ (see
Fig.~\ref{fig:Dw}). It is worth noting that the ratio between the diffusion
coefficient in seawater and that for pure water at the same temperature seems
to be independent of temperature, $D_w(S)$/$D_w(0) \approx 0.92$.
It seems interesting to check whether a Stokes-Einstein-like (SE) equation is
fulfilled in seawater using $D_w$ as a proxy for the diffusion coefficient of
the solution.
The fractional SE coefficient $t$ has been calculated from the relation
$(D/T)\propto \eta^{-t}$. The result obtained from our values in the range of
temperatures from 0 to 80\deg\ is $t=0.934$, almost the same to the value
$t=0.932$ reported for TIP4P/2005 water in the interval
280-340~K.\cite{tsimpanogiannis20}
It is interesting to note that, whereas the viscosity (and $D_w$) show a five
times increase (decrease) in the 0-80~\deg\ temperature range, the product
$\eta\;D_w$ is almost constant (Fig.~\ref{fig:Dw}.
As for the dependence of $D_w$ upon salinity, Fig.~\ref{fig:Dw} shows that
the diffusion coefficients decrease with the salt concentration. The deviation
from a linear decay is barely appreciable.
Fig.~\ref{fig:Dions} displays the self-diffusion coefficients of the ionic
components of seawater in an Arrhenius-like plot. We have also included the
rather scarce experimental measurements. The predictions for the ``in silico''
seawater model are semiquantitative. Since the simulation data seem to follow
the experimental trend for all the ions, the results for Mg$^{2+}$ are
of particular importance because the usual radioactive tracer method for the
experimental measurement of $D$ can not be applied in this case. On the other
hand, the agreement between experiment and simulation gives us some support to
extrapolate the results to higher temperatures for which no experimental data
have yet been reported. Notice finally that the self-diffusion coefficients
follow approximately an Arrhenius behavior although the fitted curves seem to
be slightly bended. 

We now turn our attention to the structure of the solution. The difficulties of
separating the contributions of each component make it extremely difficult to
obtain the distribution functions from scattering experiments. The role of
simulation in this case is complementary to experiment.

It is known that the hydration shell of the ions shows a weak dependence on
salt concentration.
Our results for seawater confirm this assertion. The position of the first
extrema of a given ion-water rdf is the same, within the
statistical uncertainty, in solutions at $S=35.165$ and $S=98.56$~g/kg. 
Only the heights of the peaks are slightly different. 
The hydration numbers are quite similar to those obtained for single salt
electrolyte solutions at very high concentrations.\cite{benavides17b,zeron19}
The HN results for the anions ---see Table~\ref{tab:lifetimes}--- seem to
increase marginally with salinity while those of the cations seem to be
independent on the salt content (the statistical uncertainty is below 0.01 for
the magnesium ion and varies from 0.01 for the more abundant ions sodium and
chloride to 0.04-0.05 for calcium and potassium).
\begin{table}
\centering
\caption{Number of water molecules in the hydration shell (HN) of the ionic
components of seawater and their average lifetimes, $\tau$ (in ps). HN is
calculated by integrating the ion-O$_w$ distribution functions up to the first
minimum, $r_{min}$ (in nm). $\tau_{integ}$ and $\tau_{decay}$ are the values of
the residence times evaluated by integration of the autocorrelation function of
HN and its fit to an exponential decay, respectively. Data correspond to
simulations at 15\deg\ and two values of the salinity.}
\label{tab:lifetimes}
\begin{tabular}{|c|rccc|rccc|}
\hline
         & \multicolumn{4}{|c}{$S$=35.165 g/kg}
                                       & \multicolumn{4}{|c|}{$S$=98.56 g/kg} \\
  Ion    &$r_{min}$&  HN &$\tau_{integ}$&$\tau_{decay}$
                                     &$r_{min}$&HN&$\tau_{integ}$&$\tau_{decay}$\\
\hline
magnesium &  0.315 &  6.00  &   -   & 1.4E5 &0.315 & 6.00  &  -   & 0.9E5 \\
 calcium  &  0.325 &  7.45  &  114  &  116  &0.322 & 7.45  & 118  &  119  \\
 sodium   &  0.315 &  5.53  &  18.2 &  18.6 &0.316 & 5.53  & 19.4 &  20.0 \\
 sulfate  &  0.468 &  12.49 &   8.5 &  9.9  &0.469 & 12.61 & 9.7  &  10.9 \\
 chloride &  0.365 &  5.86  &   7.7 &  8.2  &0.365 & 5.90  & 8.0  &  8.8  \\
potassium &  0.352 &  6.70  &   6.1 &  6.7  &0.354 & 6.67  & 6.5  &  7.3  \\
\hline
\hline
\end{tabular}
\end{table}

\begin{figure*}[tbhp]
\centering
\caption{Residence times of water molecules in the hydration shell of the ions
in seawater at 15\deg\ for two different salinities. Solid lines are the results
at $S$=35.165 g/kg and dashed lines correspond to $S$=98.56 g/kg.
In the top panels the time axis is represented on a logarithmic scale in order
to obtain a better appreciation of the behavior at short and long times for 
all the ions in the same plot.
In the bottom panels the lifetimes are represented on a logarithmic scale to
illustrate the exponential decay behavior at short times.}
\centering
\includegraphics*[clip,scale=0.34]{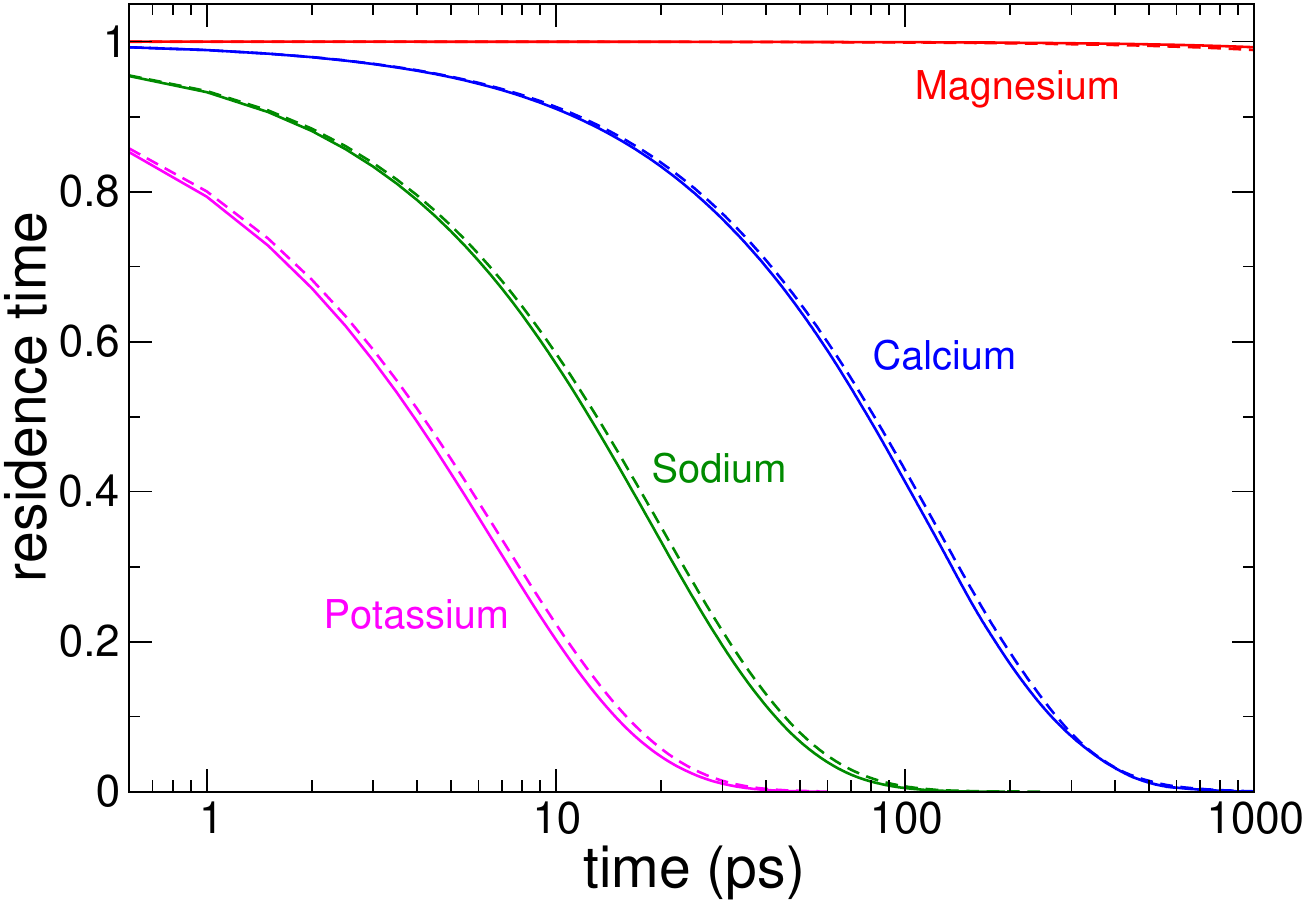}
\includegraphics*[clip,scale=0.34]{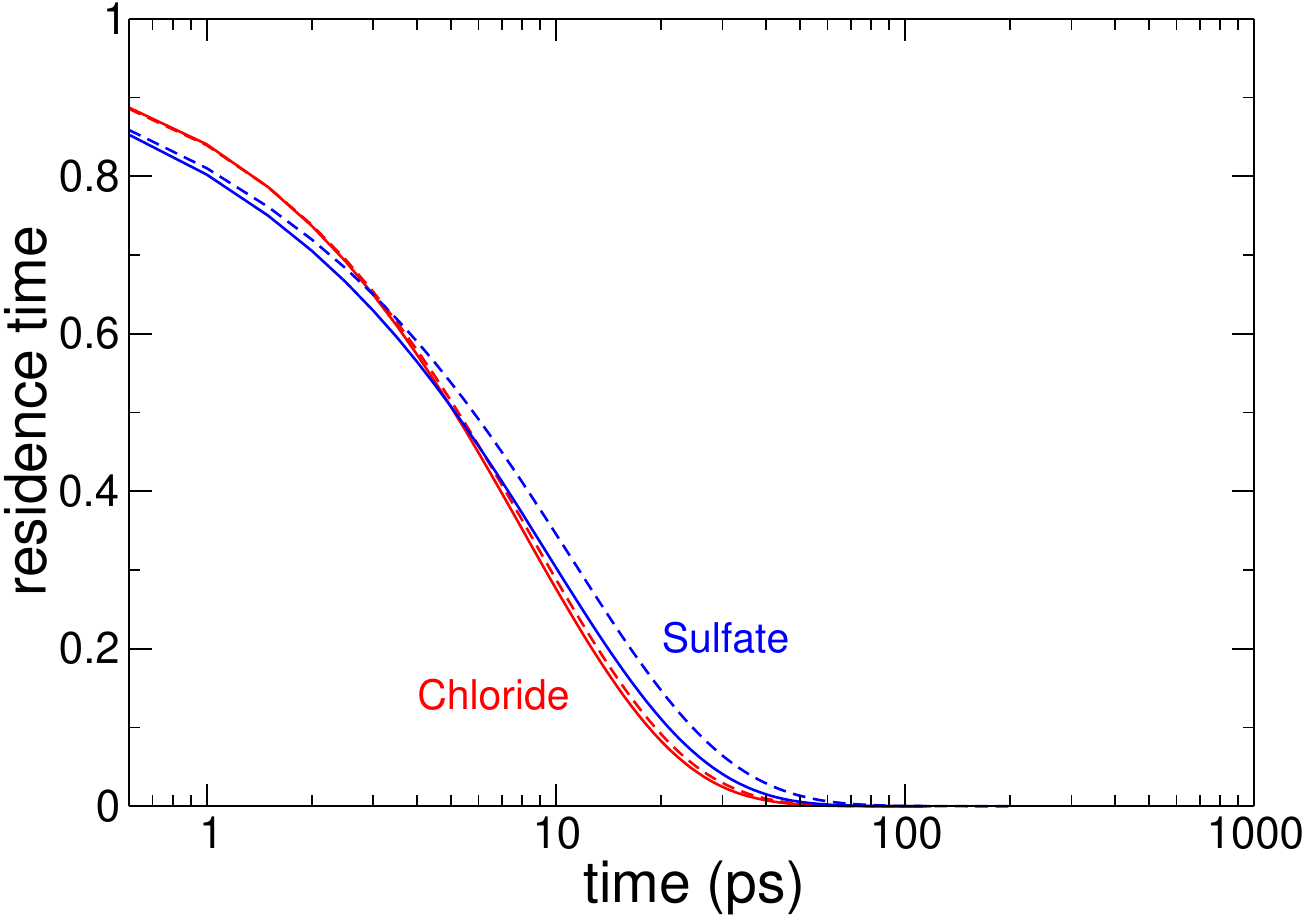} \\
\includegraphics*[clip,scale=0.29,angle=0]{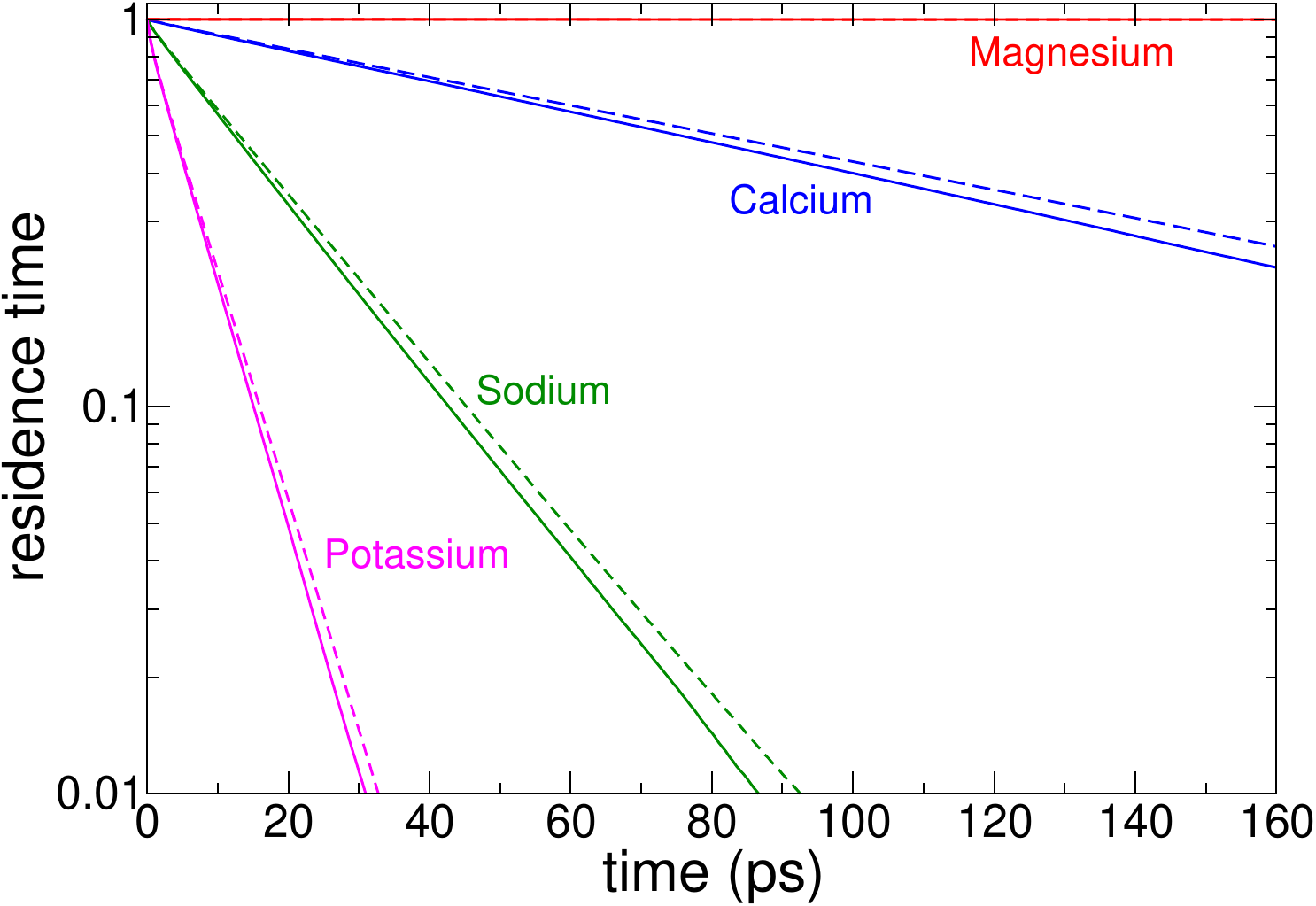}
\includegraphics*[clip,scale=0.29,angle=0]{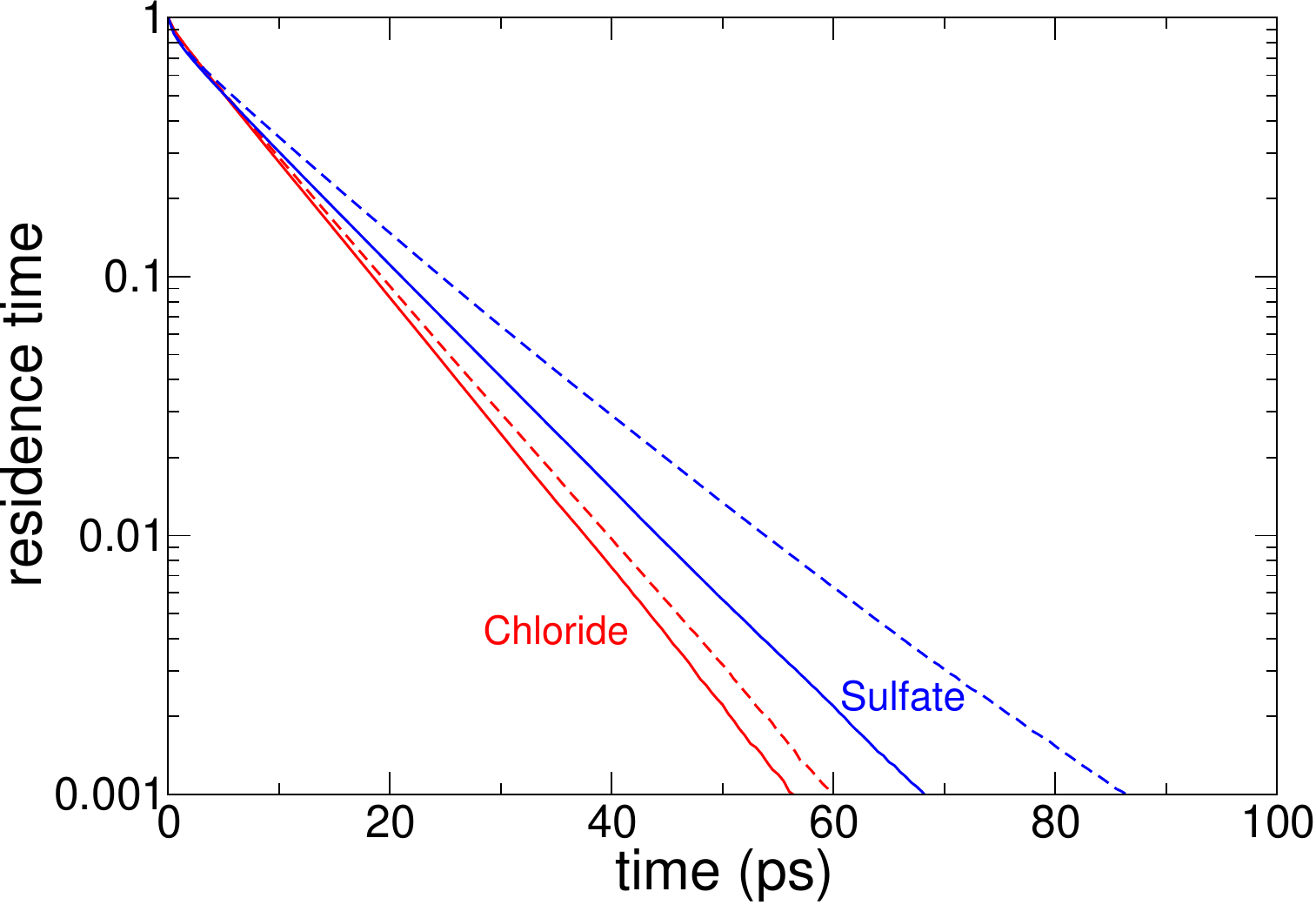}
\label{fig:lifetimes}
\end{figure*}

Little is known about the residence time, $R(t)$, of the water molecules around
ions. Our results are presented in Fig.~\ref{fig:lifetimes}.
The residence times are longer for the smaller ions. Since divalent cations are
very small, waters around them are very tightly bound. In fact, the average
lifetime of waters around magnesium ions is much longer than our simulation
length so we may only estimate it by a fit of the autocorrelation function to
an exponential decay. For cations carrying the same electronic charge, the
ionic size increases with the atomic number so the residence times for calcium
and potassium are smaller than those for magnesium and sodium, respectively. As
expected, anions exhibit a looser hydration layer than cations. The large size
of the sulfate anion motivates that, despite being a divalent ion, the lifetime
of waters around sulfates is quite similar to that of chlorides. Finally, it is
worth mentioning that the impact of a significant increase in the seawater
salinity (by a factor of about three) results in a fairly small increase in the
residence times.
	
Numerical values of the average lifetimes, $\tau$, are given in Table
\ref{tab:lifetimes}.
The values obtained using the short time exponential decay are quite similar to
those obtained by integrating the correlation function though the difference
between them is systematic. This  seems to indicate that the behavior at
longer times may deviate slightly from the exponential decay. 
The order of magnitude of the $\tau$'s is qualitatively similar to those of a
previous report for single salt aqueous solutions.\cite{koneshan98}
Quantitative differences are observed not only because of the use of
different force fields but also because of the reduced simulation lengths
in Ref.~\citenum{koneshan98}.

Given the complexity of the seawater composition, the number of radial
distribution functions (rdf) is quite large so we only summarize here the more
significant results. The presence of other ions, even if some of them are
divalent, does not affect much the sodium-chloride rdf's. We will show below
                                that the main features of the Na-Cl distribution
functions ---i.e., location and height of the maxima and minima--- are
essentially the same as those of other aqueous solutions at similar
concentrations.
The situation may be different when dealing with less abundant species,
especially when these are divalent ions.

\begin{center}
\begin{figure}[tbhp]
\caption{Mg-S(sulfate) radial distribution function (full lines) and running
coordination numbers, $n_c$ (dashed lines),
       for two solutions at 15\deg\ with the same number of water molecules,
magnesium and sulfate ions (15210, 18 and 8 respectively). One of the
systems is the ISSS+water solution at S$=35.165$~ g/kg and, thus, sodium,
chloride, calcium and potassium ions are present in the solution.
The other system does not contain those ions with the exception of 20
chloride anions required to constrain the charge balance.
The coordination number refers to the average number of magnesium cations in the
first coordination shell of a sulfate.
Notice that a different scale is used for the x-axis of the rdf (bottom) and
$n_c$ (top).}
\centering
\includegraphics*[clip,scale=0.37]{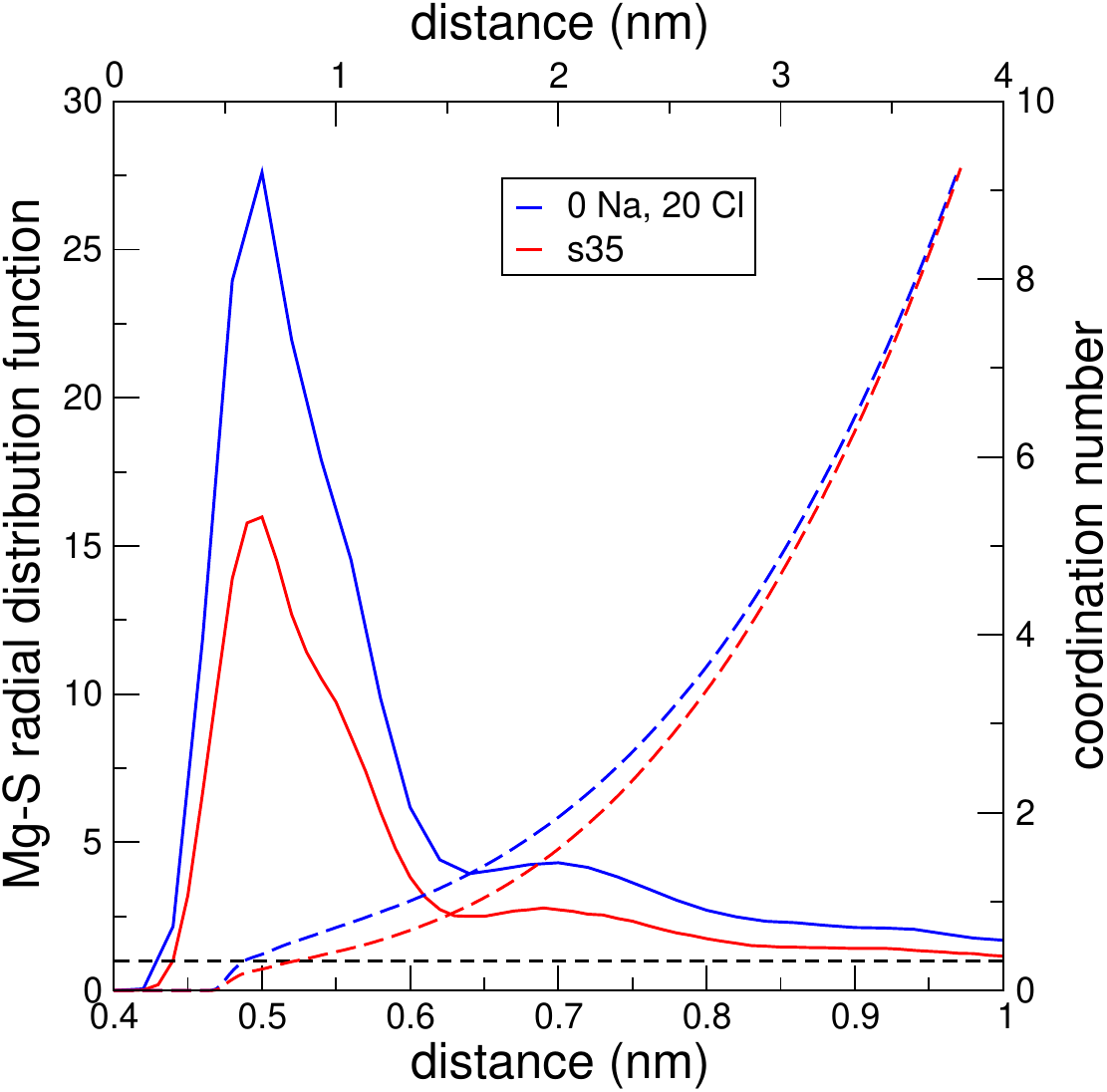}
\label{fig:rdfMgS}
\end{figure}
\end{center}

In Fig.~\ref{fig:rdfMgS} we compare the Mg-S distribution function in seawater
at $S$=35.1~g/kg to that of a solution containing only magnesium and sulfate
ions at the same concentration (though, since magnesiums outnumber sulfates
in seawater, some chloride anions must be added to enforce the charge
balance). As seen in the plot the value of the Mg-S distribution function
of the latter solution almost doubles that for seawater along a considerable
range of distances. Only at distances around 2~nm do both curves cross and the
coordination numbers of both solutions converge. This seems to be due to a
shielding effect in seawater: the divalent magnesium ions attract unlike charged
ions and, since the monovalent chloride ions are much more abundant than
sulfates, some of the chlorides surround the magnesiums thus reducing their
effective charge.

\begin{center}
\begin{figure}[tbhp]
\caption{Surface tension of seawater relative to that of pure  water.
Symbols are the results for ``in silico'' seawater using TIP4P/2005 water as
a reference; lines are experimental data\cite{nayar14,vins19}.
Results at 80\deg\ have been shifted 2~mN/m for clarity.}
\centering
\includegraphics*[clip,scale=0.55]{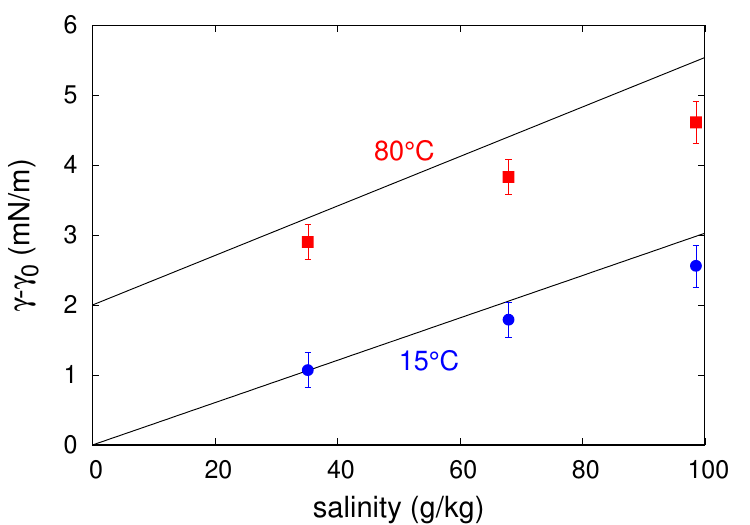}
\label{fig:surftens}
\end{figure}
\end{center}

Fig.~\ref{fig:surftens} shows the increase in the surface tension of ``in 
silico'' seawater with respect to that of pure (TIP4P/2005) water as a function
of salinity at 15 and 80\deg. The agreement with experimental results
is excellent at oceanographic conditions. Also     the variation with salinity
is almost quantitatively predicted at 15\deg. As for the 80\deg\ isotherm, the
agreement is only qualitative. The results follow the experimental trend but the
      gradient of the dependence on salinity is somewhat low, approximately 3/4
of the experimental value for $\gamma(S)-\gamma(0)$.

\subsection{Impact of modifying the number of components of the seasalt}
\label{sect:impact}
One may    wonder if the use of a more detailed
representation of the seasalt composition might affect the quality of the
results of this work. At this moment we cannot  assess directly this point
because there are no available parameters for other solutes compatible with
those of the Madrid-2019 force field. However we may obtain reasonable
suggestions.
Table~\ref{tab:vol} shows the results for the volumes calculated with the ISSS
composition compared to those obtained when replacing the interactions of the
calcium and potassium ions by those corresponding to magnesiums and sodiums,
respectively (ISSS$_{318,4}$).
Notice that the density of the solution is sensitive to the mass of the solute
components but the volume is not. Thus, a comparison of the system volumes
provides an unambiguous test of the effect of replacing the interaction
potential of a given ion by a different one.
The differences between the results at $S=35.165$~g/kg are
very small, less than 0.1~nm$^3$ though larger than the
combined statistical error (the estimated uncertainty of the results of a run
is about 0.02~nm$^3$).
The volumes for the ISSS$_{318,4}$ samples at $S=35.165$~g/kg are systematically
smaller than those of the ISSS ones but the differences are similar in the
range of temperatures investigated.
Increasing the salinity at constant temperature (15\deg) increases
the departures between both sample sets. However the differences remain quite
small, the relative deviation being less    than   0.04 percent. In summary,
the substitution of potassium and calcium ions by cations with the same
electric charge leads to an almost insignificant but detectable change of
the system volume. 
\begin{table}
\centering
\caption{Volumes (in nm$^3$) for the ISSS composition (see Table \ref{tab:ISSW})
 compared to those of solutions where the potassium and calcium ions are
 replaced by sodium and magnesium ions, respectively (ISSS$_{318,4}$).
 All the systems contain 15210 water molecules so that $S=35.165$~g/kg for the
 systems with 318 ions and six types of solutes.
 The systems at $S=67.94$~g/kg (98.56 g/kg) contain the same number of water
 molecules but duplicate (triplicate) the number of ions in the sample.}
\label{tab:vol}
\begin{tabular}{|c|c|c|c|}
\hline
t (\deg) & salinity (g/kg)&       ISSS     & ISSS$_{318,4}$  \\ \hline
  -10    &   35.165       &      458.54    &       458.46    \\ \hline
   15    &   35.165       &      459.48    &       459.43    \\ \hline
   80    &   35.165       &      473.72    &       473.66    \\ \hline
  \hline
   15    &   35.165       &      459.48    &       459.43    \\ \hline
   15    &   67.94        &      464.22    &       464.12    \\ \hline
   15    &   98.56        &      469.25    &       469.08    \\ \hline
\hline
\end{tabular}
\end{table}

The repercussion of the simplification of the ionic sample on the shear
viscosity is presented in Table~\ref{tab:viscompos}. The estimated uncertainty
of the results is less than   1.5\%. Thus, the differences between the
calculated viscosities of the ISSS and ISSS$_{318,4}$ samples are within the
combined error of both simulations.

\begin{center}
\begin{figure}[!ht]
\caption{Na-Cl radial distribution function for the ISSS and
 ISSS$_{318,4}$ compositions at s=35.165 g/kg. 
 The curve labeled as Na-Cl refers to a pure NaCl solution with the same
 number of water molecules and occupying the same volume as the
 ISSS solution.}
\includegraphics*[clip,scale=0.4]{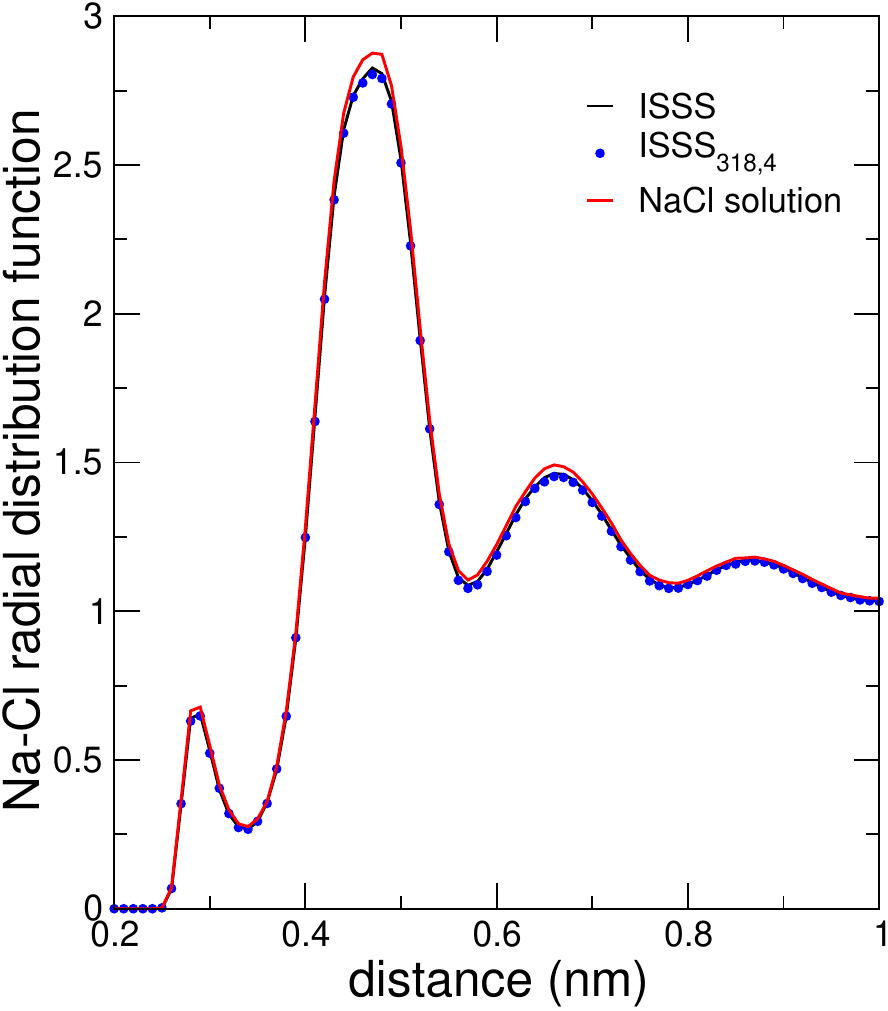}
\label{fig:rdfsISSS}
\end{figure}
\end{center}

Fig.~\ref{fig:rdfsISSS} shows a comparison of the Na-Cl distribution
functions for the ISSS and ISSS$_{318,4}$ systems at the same
temperature and salinity. 
It may be seen that the rdf is almost insensitive to the modification of
the salt composition. In fact, the difference respect to the rdf in a single
salt NaCl solution is also barely appreciable. In summary, the presence of
small amounts of additional ions has a negligible impact on the Na-Cl
distribution function.
\begin{table}
\centering
\caption{Shear viscosities (in mPa s) for the ISSS composition (see Table
 \ref{tab:ISSW}) compared to those of solutions where the potassium and calcium
 ions are replaced by sodium and magnesium ions, respectively. The systems at
 $S=17.90$ g/kg are composed of 318 ions and 30420 water
 molecules.}
\label{tab:viscompos}
\begin{tabular}{|c|c|c|c|}
\hline
t (\deg) & salinity (g/kg) &     ISSS      & ISSS$_{318,4}$  \\ \hline
   15    &   17.90        &      1.145     &       1.125     \\ \hline
   15    &   35.165       &      1.22      &       1.23      \\ \hline
   15    &   67.94        &      1.38      &       1.37      \\ \hline
   15    &   98.56        &      1.55      &       1.57      \\ \hline
  \hline
  -10    &   35.165       &      2.735     &       2.73      \\ \hline
   15    &   35.165       &      1.22      &       1.23      \\ \hline
   80    &   35.165       &      0.392     &       0.400     \\ \hline
  \hline
\end{tabular}
\end{table}

All the calculations for different properties indicate that the explicit
incorporation of the calcium and potassium ions produces marginal changes with
respect to the results obtained with a seawater sample containing only the four
most abundant ions (chloride, sodium, magnesium and sulfate).
Notice that potassium and calcium are relatively abundant components of
seawater, together they represent   1.8\% of the sea salt mole fraction.
Thus, the consequences of explicitly considering less abundant species such as the
bicarbonate and bromide anions, which represent only a 0.2\% of the sea salt,
would be much more difficult to detect in molecular simulations.
In fact, it is very likely that the impact of the incorporation of these ions
to the simulated system would be comparable or even smaller than the
uncertainty of the experimental measurements.
Further increase of the composition details beyond the ISSS composition would
only be strictly necessary if one intends to evaluate specific properties
associated to a certain ion not included in the ISSS sample.

\subsection{Comparison with the results for other force fields}
In this section we compare the results for the Madrid-2019 force field with
those obtained using an alternative set of parameters for the molecular
interactions. Since a good model for water is of paramount importance we have
also chosen TIP4P/2005\cite{abascal05b} for the alternative force field.
The more important components of seasalt are the sodium and chloride ions. The
parameters for these ions have been taken from the work of Joung and Cheatham
(JC), in particular those proposed for SPC/E water.\cite{joung08} It has been
shown that this is also an excellent choice when water is described by the
TIP4P/2005 model.\cite{dopke20}
For the sulfate anion we have selected the Cannon et al.\cite{cannon94}
potential parameters.
For the rest of the ionic solutes, we have employed the OPLS force
field\cite{jorgensen98} as implemented in the Gromacs package.
Lorentz-Berthelot combining rules were used for the cross interactions.
For simplicity we shall denote this force field as JC.
\begin{center}
\begin{figure}[!ht]
\caption{Density as a function of temperature for seawater at $S=35.165$~g/kg
using the Madrid-2019 and the JC force fields for the ions (see the text for
details about JC). TIP4P/2005 is employed for water in both cases.}
\includegraphics*[clip,scale=0.37]{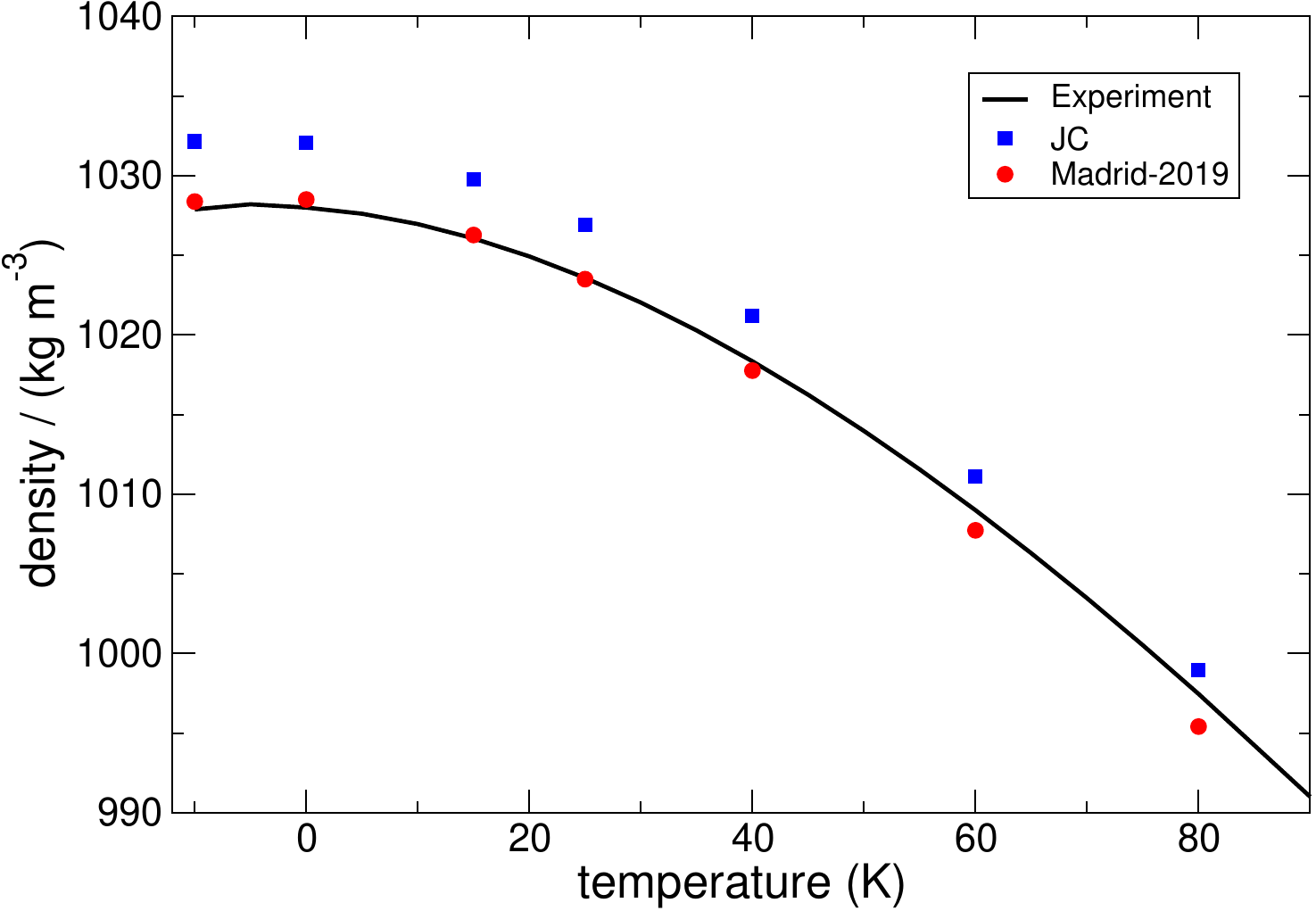}
\label{fig:dens_models}
\end{figure}
\end{center}
The densities predicted by these force fields for a salinity $S=35.165$~g/kg
are shown in Fig~\ref{fig:dens_models}. Overall the Madrid-2019 model
performs better than JC.
The predictions of the JC force field are noticeably different from the
experimental densities at low to medium temperatures but the discrepancies
decrease with increasing temperatures. Eventually, at 80\deg, the performance
of the JC model is slightly better than that of the Madrid-2019 force field.

\begin{center}
\begin{figure}[!ht]
\caption{Shear viscosity as a function of temperature for seawater at $S=35.165$
g/kg using the Madrid-2019 and the JC force fields for the ions and TIP4P/2005
for water (see the text for details.)}
\includegraphics*[clip,scale=0.70]{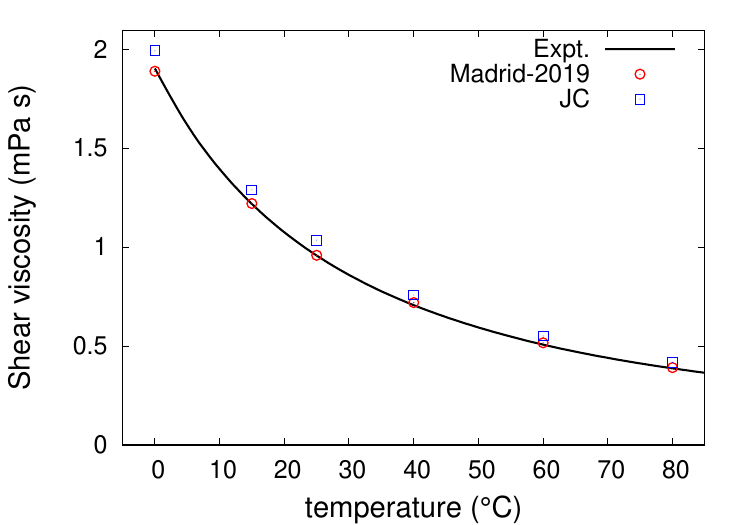}
\label{fig:viscomodels}
\end{figure}
\end{center}

As for the shear viscosity, we observed the formation of calcium sulfate
aggregates in the JC simulations. This may be expected considering that the
concentrations of these ions in seawater are not far from the solubility limit
of the salt. Increasing the Lorentz-Berthelot rules by a factor of 1.1 solved
this issue (densities shown in Figure 9 were also obtained with this prescription).
Fig.~\ref{fig:viscomodels} indicates that the performance of the Madrid-2019
force field is much better than that of the JC model at all the temperatures
investigated for $S=35.165$~g/kg.

\section{Conclusions and final discussion}
We have shown in this work that molecular dynamics simulations of a solution
mimicking the composition of seawater using a state of the art force field
yield results in excellent agreement with experimental data. The quality of the
predictions is
particularly impressive in the oceanographic range but it is also very good for
conditions relevant to desalination processes. It is difficult to assess at
this moment the importance of this fact. Notice that the outcome of the
simulation not only is a wide set of macroscopic properties but also
gives detailed microscopic information which is sometimes very difficult to
obtain   in experiments. In this work we have calculated relevant magnitudes
for which there is a large set of experimental determinations. It is the case
of the density, viscosity or surface tension. The quality of the predictions
validate the force field employed and our simplified representation of the
seawater composition. This gives support to the predictions for structural
quantities such as the hydration numbers and residence times of water molecules
around ions as well as the ion-ion distribution functions. We have also
obtained reliable results for the ionic diffusivity, a magnitude for which the
experimental data are extremely scarce.

Once the force field has been verified, molecular simulation also allows to
carry out ``what if?'' pseudo experiments. One may investigate the impact of
small changes in composition by modifying the amount of any constituent.
In fact, we have reported in this work the effect of replacing the calcium and
potassium ions by magnesium and sodiums, respectively. Although we have limited
our study to a seasalt containing only six types of ions nothing prevents us, in
principle, from extending our investigations to a more detailed composition. For
instance, it would be very interesting to learn something about the behavior of
the components involved in CO$_{\rm 2}$ sink in oceans\cite{feely04} and
CO$_{\rm 2}$ sequestration in deep saline aquifers\cite{celia15,shi18}.
The diffusion coefficient of CO$_{\rm 2}$ in brines has already been
investigated by molecular simulation(see for instance Ref.
\citenum{garciarates12}). However, similarly
to the case of the salting out effect of methane\cite{blazquez20} we have
preliminary results indicating that the solubility of CO2 in a NaCl solution
may not be described correctly by most of the current force fields. In
addition, it would be necessary to fit the interaction parameters of CO2 with
the main ionic components of seawater beyond standard LB rules.
Although the design of appropriate force fields for them is not a trivial task,
it is nowadays within our reach. In fact, we hope that this study will
stimulate work in that direction.

\begin{acknowledgement}
This work has been funded by Grants
PID2019-105898GB-C21 (Ministerio de Educaci\'on, Spain) and GR-910570 (UCM).
I.M.Z. thanks CONACYT (M\'exico) for the financial support (Estancias
Posdoctorales en el Extranjero) and M.A.G. acknowledges funding by
Ministerio de Educacion, Spain (Juan de la Cierva fellowship, IJCI-2016-27497).
\end{acknowledgement}

\begin{suppinfo}

A description of the procedure to obtain optimal samples for the simulation of
seawater. These samples reproduce as close as possible the Reference
Composition with a minimum of solute molecules for a given number of components.
The document also includes a detailed table of the number of solute molecules in
several optimal samples for up to 13 seasalt components.
\end{suppinfo}

\end{document}